\documentclass[11pt,a4paper]{article}
\usepackage{jcappub}
\usepackage{float}
\usepackage{rotating}
\usepackage{graphicx}
\usepackage{epsfig}
\usepackage{amsmath}
\usepackage{amssymb}
\usepackage{mathrsfs}  
\usepackage{subfigure}
\usepackage{time}
\usepackage{relsize, comment}
\usepackage{xcolor}
\usepackage[utf8]{inputenc}

\hypersetup{ linktoc=all,
    colorlinks=true, linkcolor={blue},  
       citecolor={red}, urlcolor={vividviolet}
}
\definecolor{rosy}{RGB}{230,235,252}
\definecolor{myframetitle}{RGB}{90,89,170}
\definecolor{myblocktitle}{RGB}{140,185,249}
\definecolor{mytitle}{RGB}{10,80,26}

\definecolor{darkgreen}{RGB}{27,130,45}
\definecolor{darkblue}{rgb}{0,0,0.3}
\definecolor{darkred}{rgb}{0.7,0,0}

\definecolor{light gray}{RGB}{220,220,220}
\definecolor{dark purple}{RGB}{108,0,217}
\definecolor{pink}{RGB}{190,20,100}
\definecolor{orang}{RGB}{193,63,0}
\definecolor{green}{RGB}{11,98,17}
\definecolor{darkpink}{RGB}{153,0,76}
\definecolor{bluegreen}{RGB}{0,102,102}
\definecolor{greenlagan}{RGB}{0,102,0}
\definecolor{redgreen}{RGB}{102,102,0}
\definecolor{Redgreen}{RGB}{153,76,0}
\definecolor{vividviolet}{rgb}{0.62, 0.0, 1.0}
\definecolor{amaranth}{rgb}{0.9, 0.17, 0.31}
\definecolor{palatinateblue}{rgb}{0.15, 0.23, 0.89}
\definecolor{brightpink}{rgb}{1.0, 0.0, 0.5}
\definecolor{cornflowerblue}{rgb}{0.39, 0.58, 0.93}
\definecolor{deepcarminepink}{rgb}{0.94, 0.19, 0.22}
\definecolor{radicalred}{rgb}{1.0, 0.21, 0.37}
\title{Testing No slip model with pulsar timing arrays: NANOGrav and IPTA}

\author[a]{Mohammadreza Davari}
\emailAdd{m.davari@khu.ac.ir}
\author[a,1]{Alireza Allahyari \note{corrsponding author}
}
\emailAdd{alireza.al@khu.ac.ir}
\author[a]{Shahram Khosravi}
\emailAdd{khosravi\_sh@khu.ac.ir}
\affiliation[a]{Department of Astronomy and High Energy Physics, Kharazmi University, 15719-14911, Tehran, Iran \looseness=-1}

\abstract{We perform an observational study of modified gravity considering a potential inflationary interpretation of  pulsar timing arrays (PTA). We use a motivated model known as no slip in which the gravitational wave propagation is modified. Specifically, by using  two different parametrizations for the model, we find the approximate transfer functions for tensor perturbations. In this way, we obtain the spectral energy density of gravitational waves and use NANOGrav and IPTA second data release to constrain parameters of the model. We find that there is  degeneracy between the model parameters $\xi$ and $c_M$. For $c_M$, we only get an upper bound on the parameter. Thus, it is difficult to constrain them with percent level accuracy with the current  PTA data. }

\begin{document}
\maketitle
\flushbottom
\section{Introduction}
Generation of the stochastic  gravitational waves background are predicted by various astronomical and early universe models.
Recently, there are also  hints towards their existence by various  observations like NANOGrav \cite{NANOGrav:2023gor}, Parkers PTA \cite{Reardon:2023gzh}, European PTA \cite{EPTA:2023fyk} and the China PTA \cite{Xu:2023wog}. There are different mechanisms which could generate a stochastic background. Two prevailing views for the sources of these gravitational waves (GW) \cite{NANOGrav:2023hvm} are the superpositions of GW signals from the merging of supermassive black hole binaries \cite{NANOGrav:2023gor,NANOGrav:2023pdq,Burke-Spolaor:2018bvk,Aggarwal:2018mgp,EPTA:2015gke,Mingarelli:2017fbe,Bian:2020urb,McWilliams:2012an,Sesana:2017lnk,Chen:2018znx,Middleton:2017nbg,Sato-Polito:2023spo,Yunes:2016jcc} and the primordial GWs from early universe \cite{Caprini:2018mtu,Madge:2023dxc,Ellis:2023oxs,Wu:2023hsa,Benetti:2021uea}. Other mechanisms which could contribute to this background include induced gravitational waves from scalar perturbations \cite{Guzzetti:2016mkm,Domenech:2021ztg,Vagnozzi:2020gtf,Vagnozzi:2023lwo,Yi:2023mbm,Unal:2023srk,Servant:2023mwt,Lu:2023mcz,Borah:2023sbc,Franciolini:2023pbf}, first order cosmological phase transitions \cite{Kosowsky:1992rz, Kamionkowski:1993fg, Caprini:2007xq, Hindmarsh:2013xza,Megias:2023kiy} and domain walls or cosmic strings \cite{Kibble:1976sj,Vilenkin:1981bx,Caldwell:1991jj,Vilenkin:1981zs,EuropeanPulsarTimingArray:2023lqe,Ellis:2020ena,Sanidas:2012ee,Ellis:2023tsl,Buchmuller:2021mbb}. Focusing for concreteness on the NANOGrav signal, this stochastic GW background has a blue-tilted tensor spectrum, with the spectral index $n_T = 1.8 \pm 0.3$ \cite{Vagnozzi:2023lwo}.   The primordial production of GW background is possible as there are some mechanisms to generate blue-tilted power spectrum, for instance within “phantom” inflationary models \cite{Piao:2004tq}. In these models $\omega < -1$ and  one expects $n_T > 0$. Thereby, a blue-tilted spectrum is achieved \cite{Vagnozzi:2023lwo}.  There are also mechanisms which produce a modified power law spectrum \cite{Benetti:2021uea,Jiang:2023gfe,Kuroyanagi:2018csn}. The primordial origin for the GW background is investigated in \cite{Ben-Dayan:2023lwd, Choudhury:2023kam, Kuroyanagi:2020sfw}.

PTAs provide an arena to  look for modifications to gravity. The study of GWs in  modifications of gravity has also been the subject of focused studies \cite{Liang:2023ary,Nunes:2018zot,Arai:2017hxj,Lin:2016gve,Nishizawa:2017nef,Belgacem:2018lbp,Alves:2016iks,Nunes:2018evm,Casalino:2018wnc,Visinelli:2017bny,Fujita:2018ehq,Lombriser:2015sxa,BeltranJimenez:2018ymu,Isi:2018miq,Cornish:2017oic,Scomparin:2019ziw,Kramer:2006nb,Freire:2012mg,Tahara:2020fmn}.
The spectrum of primordial GWs is not only determined by the evolution of the background cosmology, but also it can be significantly affected by modifications to  General Relativity. 

The Horndeski theories of gravity \cite{Horndeski:1974wa,Deffayet:2011gz,Kobayashi:2011nu,CANTATA:2021ktz,Tsujikawa:2014mba} are the most general scalar-tensor theories with second-order equations of motion. Modified gravities predict  different generation and propagation mechanisms for GWs. Additionally, we need to consider the fact that the recent measurement of the speed of propagation of GWs from GW170817 relative to its electromagnetic counterpart GRB170817A \cite{LIGOScientific:2017zic} severely limits the deviations from the speed of light. In the  subclass of Horndeski theories, a motivated model named no slip model does not alter the speed of GWs. Moreover, the slip between metric potentials, should vanish and yet the gravity theory does not reduce to general relativity \cite{Linder:2018jil}. 

In this paper, we will constrain the parameters of no slip gravity model as a subclass of Horndeski theories, using NANOGrav 15-year data set (NG15) \cite{NANOGrav:2023gor} and International PTA second data release (IPTA2) \cite{Perera:2019sca,Antoniadis:2022pcn}.
We assume that GW background has a primordial origin from an inflationary era and after the inflationary era, universe is described by no slip model \cite{Ben-Dayan:2023lwd}.
We study two different parametrizations of the model, taking a phenomenological approach. In both parametrizations, we assume one extra parameter to constrain. We compare our results with  our previous study using the standard sirens approach combined with cosmic microwave background.  For various recent works on GWs in modified gravity models as an explanation of the PTA observations see \cite{Domenech:2024drm,Bernardo:2023pwt,Fu:2023aab,Tasinato:2023zcg,Higashino:2022izi,Zhang:2023lzt,Boitier:2020xfx,Choudhury:2024one,Hu:2023vsg,Wu:2023rib,Cannizzaro:2023mgc,Dong:2022zvh,ElBourakadi:2022anr}.

This paper is organized as follows. Section ~(\ref{sec2}) is devoted to no slip model. We adopt  two   phenomenological parametrizations. Additionally, we consider the propagation of GWs in no slip model. 
In Section~(\ref{sec3}), we  derive the approximate  transfer functions for tensor perturbations for this model and find the spectral energy density.  By using the NANOGrav and IPTA data, we constrain the model. Section~(\ref{sec5}) is summary and discussion.

\section{No slip model} \label{sec2}
Let us assume a flat FRLW background.
The scalar and tensor perturbations on this background are   are given by
	\begin{equation}
		ds^2=a(\eta)^2\left\lbrace -\left(1+2 \Psi \right)d\eta^2+\left(1-2\Phi \right)\delta_{ij}dx^{i}dx^j+h_{ij} dx^{i}dx^j  \right\rbrace \,,
	\end{equation}
 where $\Phi$ and $\Psi$ are scalar perturbations in conformal Newtonian gauge and $h_{ij}$ denotes tensor perturbations.
	In no slip model, the effects of gravity on observations of matter and light in the universe can  be suitably described by modified Poisson equations relating the time-time metric potential $\Psi$ and space-space metric potential $\Phi$ (in Newtonian gauge) to the matter perturbations  as \cite{Linder:2018jil}
	\begin{equation}
	\nabla^2 \Psi = 4 \pi G_N \delta\rho \times G_{matter}\,, \quad \quad
	\nabla^2 (\Psi+\Phi) = 8 \pi G_N \delta\rho \times G_{light}\,,
	\end{equation}
	where the first equation governs the growth of structures, with a gravitational strength $G_{matter}$, and the second governs the deflection of light, with a gravitational strength $G_{light}$. The offset between $G_{matter}$ and $G_{light}$, or $\Psi$ and $\Phi$, is referred to as the gravitational slip \cite{Linder:2018jil} with
	\begin{equation}
	\bar\eta \equiv \frac{G_{matter}}{G_{light}}\,,
	\end{equation}
    where $\bar\eta = 1$ corresponds to vanishing slip. The expressions for $G_{matter}$, $G_{light}$, and slip in Horndeski gravity, or the equivalent effective field theory
    (EFT) approach, are given in \cite{Bellini:2014fua,Gubitosi:2012hu,Linder:2015rcz}.
\subsection{Modified gravitational wave propagation}

	The evolution of linear, transverse-traceless perturbations for the tensor modes due to modifications of the gravity theory are generally described by the following equation \cite{Saltas:2014dha}
	\begin{equation}
		\ddot{h}_{ij} + (3 + \nu) H \dot{h}_{ij} + ( c_T^2 k^2/a^2 + \mu^2)h_{ij} = \Gamma_{ij}\,,
	\end{equation}
	where $h_{ij}$ is the metric tensor perturbation and dot denotes derivative with respect to comoving time $dt=a(\eta) d\eta$. The four time dependent parameters are as follows. $c_T$ is the GW propagation speed, $\mu$ is the effective graviton mass, $\nu$ is related to the running of the effective Planck mass, and $\Gamma_{ij}$ denotes the anisotropic stress generating GWs. Assuming there is  no anisotropy  and adapting the above equation in the context of the Horndeski gravity, we have \cite{Nunes:2018zot}
	\begin{equation}
		\ddot{h}_{ij} + (3+ \alpha_{M})H \dot{h}_{ij} + (1+ \alpha_{T})\frac{k^2}{a^2} h_{ij} = 0\,,
	\end{equation}
	where we have identified $\nu = \alpha_M$ , $c^2_T = 1 + \alpha_T$ , $\mu = 0$, and also $\alpha_M$ and $\alpha_T$ are two dimensionless functions given by \cite{Nunes:2018zot}.  We have
	\begin{equation}
		\alpha_M = \frac{1}{HM_*^2} \frac{dM_*^2}{dt}\,,
	\end{equation}

	the definitions for $M_*$ 
	can be found in \cite{Nunes:2018zot}.
	The no slip model has the property that the speed of propagation of gravitational waves
	equals to the speed of light, that is $c^2_T = 1$ or $\alpha_T = c^2_T - 1 = 0$ \cite{Linder:2018jil}; So we have
	\begin{equation}
		\ddot{h}_{ij} + (3+ \alpha_{M})H \dot{h}_{ij} + \frac{k^2}{a^2} h_{ij} = 0\,.
	\end{equation}
	Also comparing with the equation in \cite{Allahyari:2021enz}, we have
	 $q_t = 2G_4 = M_\star^2$ where the definition for $G_{4}$ is given in \cite{Nunes:2018zot}. We take a phenomenological approach  and parameterize the theory in terms of well motivated parameters \cite{Allahyari:2021enz, Linder:2018jil} used in a standard siren study of GWs with LISA and cosmic microwave background study \cite{Linder:2018jil}. In this work, we apply these two common parameterizations for no slip model:
	
	\textbf{Parameterization I}: According to the parametrization introduced in \cite{LISACosmologyWorkingGroup:2019mwx} for no slip model, the parameter $\alpha_M$ can be expressed as
	\begin{equation}
		\alpha_{M} = -2\delta(z) = \frac{d\ln q_t}{d\ln a} = \frac{-2n(1-\xi)}{1-\xi+\xi(1+z)^n}\,,
	\end{equation}
	where $\xi$ and $n$ are constants and $z$ is the redshift. More specifically, to relate these parameters with $M_*$, we have \cite{Mitra:2020vzq}
	\begin{align}
		\xi =  \lim_{z\to \infty}\frac{M_{*}(0)}{M_{*}(z)}\,,\\
		n \simeq \frac{\alpha_{M0}}{2(\xi-1)}\,,
	\end{align}
	with $\alpha_{M0} = -2\delta(0)$. This parametrization is applicable to models in which we expect some simple properties in low and high redshifts. More explicitly, we find that at the high redshifts $z \rightarrow \infty$, we have $\delta(z) \rightarrow 0$. Moreover, when $z \rightarrow 0$, this yields $\delta(z) \rightarrow n(1-\xi)$, where for $\xi = 1$ we have $\delta(z) \rightarrow 0$. Thus, we recover the standard expression for the luminosity distance at high and low redshifts \cite{Allahyari:2021enz}.
	
	\textbf{Parameterization II}: According to the parametrization introduced in \cite{Linder:2018jil} and used in \cite{Brush:2018dhg} to investigate the effects of no slip gravity for CMB lensing and B-modes; we have
	\begin{equation}
		\alpha_{M} = \frac{4c_M(a/a_t)^\tau}{[(a/a_t)^\tau +1]^2}\,,
	\end{equation}
	where $c_M$ is the amplitude of the running of the Planck mass, $a_t$ is the scale factor at the transition time and $\tau$ is the rapidity. For this form the stability condition requires $c_M\geq0$ and $0<\tau \leq3/2$ \cite{Linder:2018jil}.

\section{Primordial GWs and PTAs} \label{sec3}
In this section, we derive the expressions for the spectral energy density $\Omega_{GW}$. Moreover, we find tensorial transfer function and relate $\Omega_{GW}$  to primordial power spectrum. We use this expressions to find constraints on the relevant parameters.
 
  The relevant quantity in PTA observations is the spectral  energy density.
  The spectral energy density of GWs can be derived as  \cite{Caprini:2018mtu}
	\begin{equation}
		\label{spec}
		\Omega_{GW}(k,\eta) = \frac{1}{12H^2a^2}[T'(k,\eta)]^2 P_t(k)\,,
	\end{equation}

	where $T(k,\eta)$ is the transfer function. Transfer function describes the  evolution of GW modes after the modes re-enter the horizon. The quantity $P_t(k)$ is the primordial power spectrum of GWs at the end of the inflationary period. We may write this in terms of a tilt $n_t$ and a tensor amplitude $A$ as
	
	\begin{equation}
	P_t(k) = A(\frac{k}{k_{\star}})^{n_t}\,,
	\end{equation}
	where 	$k_{\star}$ is a pivot scale set as $k_{\star}=0.05\,{\rm Mpc}^{-1}$.
	 The transfer function in modified gravity  can be approximated as a product of a correction factor and a general relativity part. For no slip model we have \cite{Nunes:2018zot}
	\begin{equation}
		\label{transferMG}
		T(k,\eta)_{MG} = \mathrm{exp}\left(-\frac{1}{2} \int_{}^{\eta} \alpha_M \mathcal{H} \mathrm{d}\eta'\right) T(k,\eta)_{GR}=e^\mathcal{D}\,T(k,\eta)_{GR}\,,
	\end{equation}
    where $\mathcal{H}$ is the Hubble parameter in terms of conformal time. The exponential term in equation \ref{transferMG} is the correction factor.

    We plot this factor as a function of redshift using the consistent values  for the parameters  given in section~\ref{sec3}
    from NG15 in our analysis. This factor reaches unity asymptotically in higher redshifts.
    \begin{figure}[htbp]
    	\centerline{\subfigure{\includegraphics[scale=0.4]{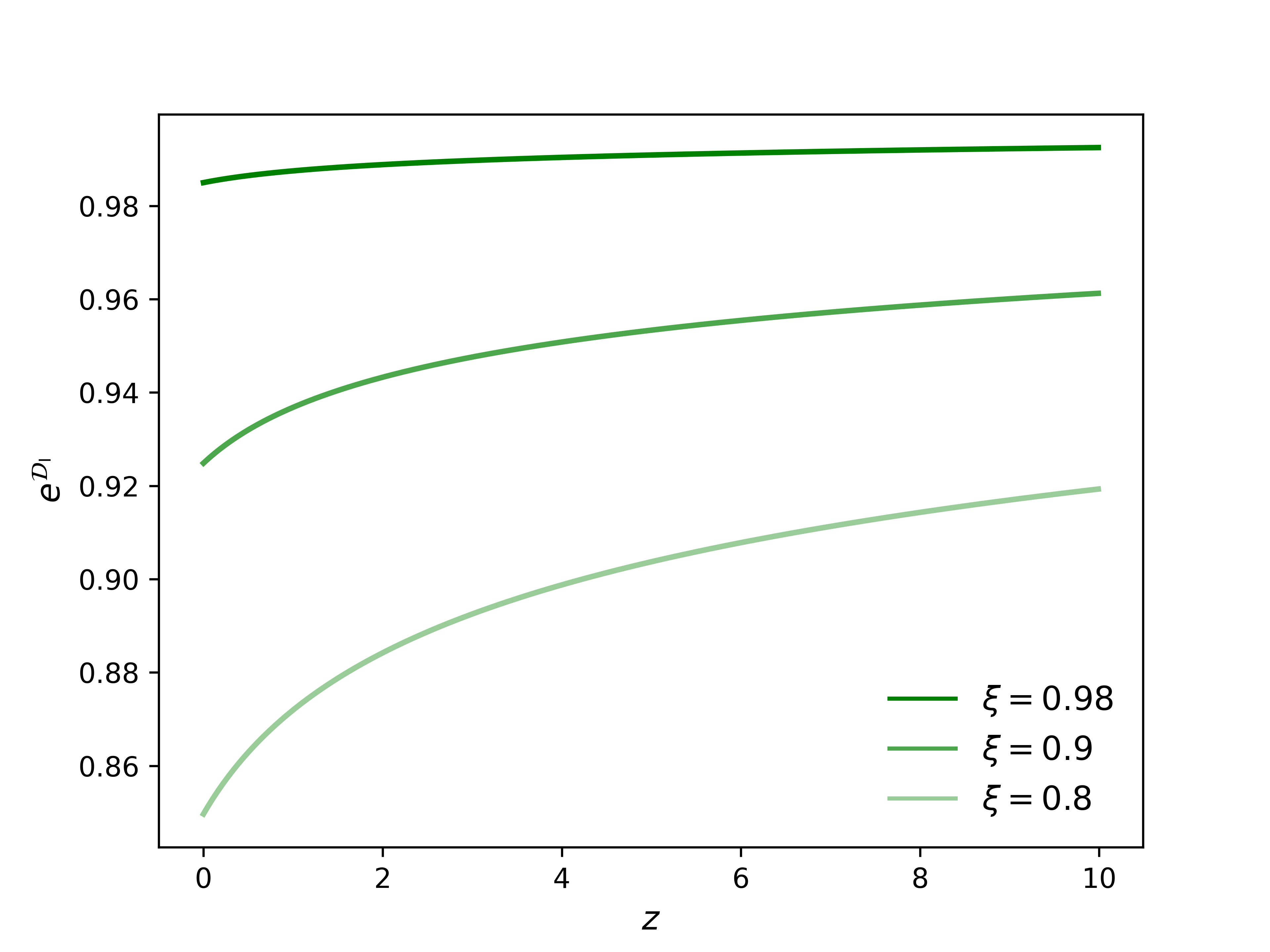}}
    		\subfigure{\includegraphics[scale=0.4]{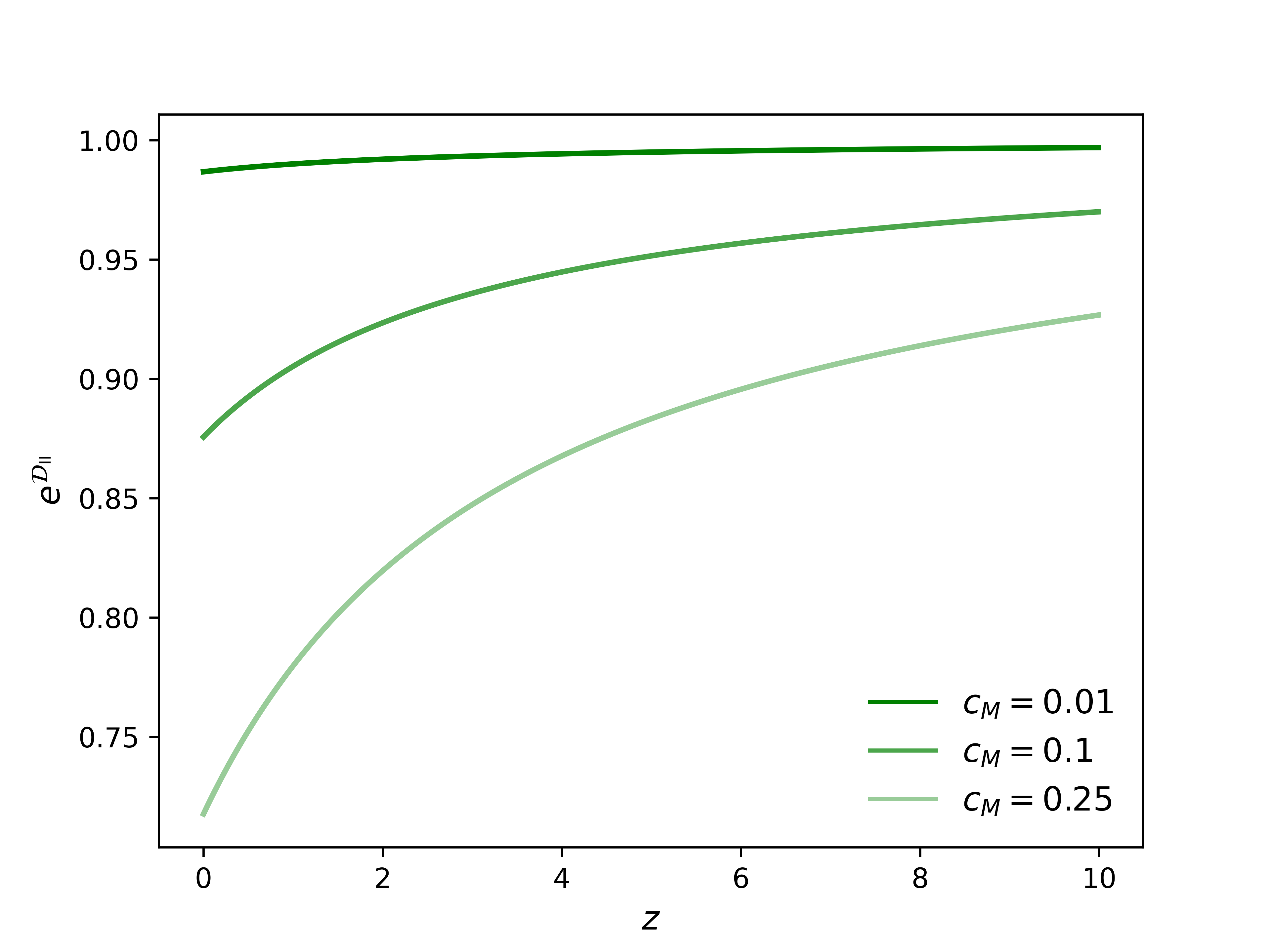}}}
    	\caption{The correction factor  in terms of redshift z. \textbf{Left}: parametrization I. \textbf{Right}: parametrization II. }
    	\label{figgrc}
    \end{figure}

    The transfer function and its time derivative in context of GR, are numerically computed in \cite{Zhao:2006mm, Zhao:2013bba} and  are given by
    \begin{equation}
    	T(k,\eta_0)_{GR} = \frac{3j_1(k\eta_0)}{k\eta_0} \sqrt{1.0 + 1.36(\frac{k}{k_{eq}}) +2.50(\frac{k}{k_{eq}})^2}\,,
    \end{equation}
    \begin{equation}
    	T'(k, \eta_0)_{GR} = \frac{-3j_2(k\eta_0) \Omega_m}{k\eta_0} \sqrt{1.0 + 1.36(\frac{k}{k_{eq}}) +2.50(\frac{k}{k_{eq}})^2}\,,
    \end{equation}
    where $j_1$ , $j_2$ are Bessel functions and $\eta_0$ is the present conformal time. For GWs in PTA scales, we can use the approximation $(k \gg k_{eq})$ \cite{Vagnozzi:2023lwo}. Then from equations \ref{transferMG},  the transfer functions for both parametrizations is obtained as
    \begin{equation}
    	T(k,\eta)_{\text{I}} = \left(\frac{a^n (1- \xi) +\xi}{\xi}\right) \ T(k,\eta)_{GR}\,,
    \end{equation}    
    \begin{equation}
    	T(k,\eta)_{\text{II}} = \mathrm{exp}\left(\frac{2c_M}{\tau}\left[\frac{a^\tau + a_t^\tau}{a_t^\tau (1+(\frac{a}{a_t})^\tau)^2}-1\right]\right) \ T(k,\eta)_{GR}\,,
    \end{equation}   
    where we assumed $n>0$ and $\tau>0$ for integration. Subscript I (II) denotes the parametrization I (II), respectively.

    In pulsar timing experiments, it is convenient to express wavenumbers $k$ in terms of frequencies $f$ as \cite{Vagnozzi:2023lwo}  
    \begin{eqnarray}
     f \simeq 1.54 \times 10^{-15} \left ( \frac{k}{{\rm Mpc}^{-1}} \right ) \,{\rm Hz}\,.
    \label{ktof}
    \end{eqnarray}
    We find that
    $f_{\star} \simeq 7.7 \times 10^{-17}\,{\rm Hz}$. Also in PTAs, the GW spectral energy density is rather written in terms of the power spectrum of the GW strain $h_c$ given by
    \begin{eqnarray}
    \Omega^{\rm PTA}_{\rm gw}(f) = \frac{2\pi^2}{3H_0^2}f^2h_c^2(f)\,.
    \label{eq:omegagwhc}
    \end{eqnarray}
    $h_c(f)$ is supposed to take a powerlaw form with respect to a reference frequency $f_{\rm yr}$ and can be expressed as
    \begin{eqnarray}
    h_c(f) = A \left ( \frac{f}{f_{\rm yr}} \right ) ^{\alpha}\,,
    \label{eq:hc}
    \end{eqnarray}
    where $f_{\rm yr}=1\,{\rm yr}^{-1} \approx 3.17 \times 10^{-8}\,{\rm Hz}$.
    
    Finally  using $\alpha = \frac{3-\gamma}{2}$,
    the present GWs spectral energy density  can be found \cite{Vagnozzi:2023lwo}. We have the spectral energy density of GWs as
    \begin{equation}\label{omegaI}
    	\Omega_{GW}(f)_{\text{I}} = A^2 \frac{2\pi^2}{3H_0^2} \left(\frac{1}{\xi} + \frac{nH_0(1-\xi)}{2\pi f \xi}\right)^2  \left(\frac{f^{5-\gamma}}{yr^{\gamma-3}}\right)\,,
    \end{equation}
    \begin{equation}\label{omegaII}
    	\Omega_{GW}(f)_{\text{II}} = A^2 \frac{2\pi^2}{3H_0^2} \mathrm{exp}\left(\frac{4c_M}{\tau}\left[\frac{1}{ (1+(\frac{1}{a_t})^\tau)^2}-1\right]\right) \left(\frac{-c_M H_0}{\pi f a_t^{\tau}(1+ (\frac{1}{a_t})^{\tau})^2} + 1\right)^2  \left(\frac{f^{5-\gamma}}{yr^{\gamma-3}}\right)\,,
    \end{equation}
    \begin{equation}
        \gamma = 5 - n_t\,.\notag
    \end{equation}    
    where we consider $H_0 = 67.36$ from Planck 2018 \cite{Planck:2018vyg}.
   \subsection{Constraints by PTAs}
    We set to constrain the parameters of theory by PTA observations. We use NANOGrav 15-year data (NG15) and IPTA second data release (IPTA2).
    To constrain the parameters $\xi$ and $c_M$, we use the python package  \texttt{PTArcade} \cite{Mitridate:2023oar}, that is a wrapper of \texttt{ENTERPRISE} \cite{ellis2019enterprise} and \texttt{ceffyl} \cite{Lamb:2023jls} to provide an accessible way to perform Bayesian analyses  with PTA data. In this work we use \texttt{ceffyl} with NG15 and IPTA2 and use uniform priors on the parameters as $(-18<log_{10}A<-6)$, $(0<\gamma<6)$ for both parameterizations.  For \emph{parameterization I}, we set $(n=0.2)$ from \cite{Allahyari:2021enz} and $(0<\xi<2)$; and for \emph{parameterization II} we set $(\tau=1)$ from \cite{Brush:2018dhg} and $(0<c_M<0.5)$.
    
    \begin{figure}[htbp]
    	\centerline{\includegraphics[scale=1.0]{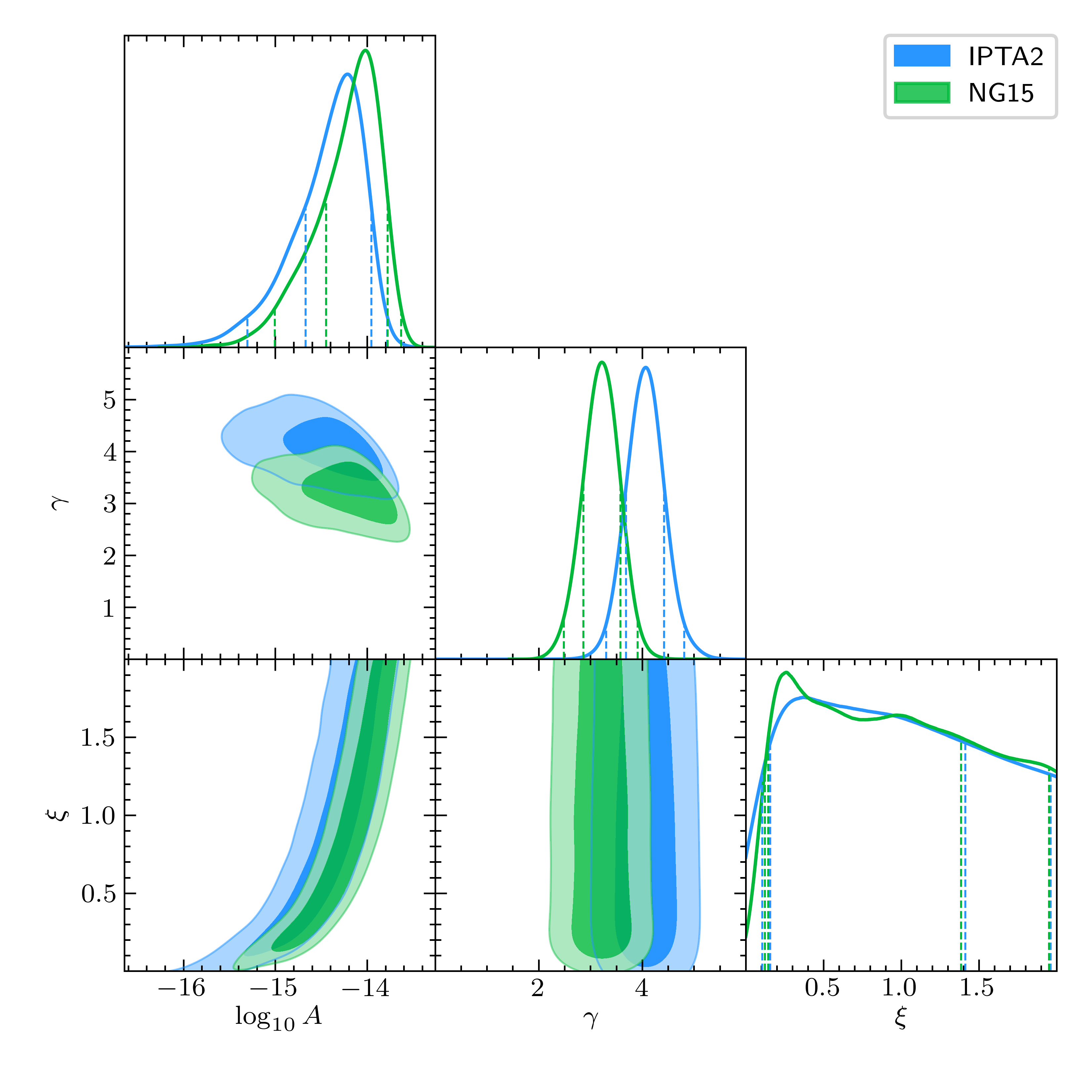}}
    	\caption{The posterior plots and marginal posteriors of no slip model parameters for \emph{parameterization I}. The contours show at 68\% and 95\%  confidence levels for International PTA second data release (IPTA2) in blue and NANOGrav 15-year data set (NG15) in green.}
    	\label{figfppar1}
    \end{figure}
    \begin{figure}[htbp]
    	\centerline{\includegraphics[scale=1.0]{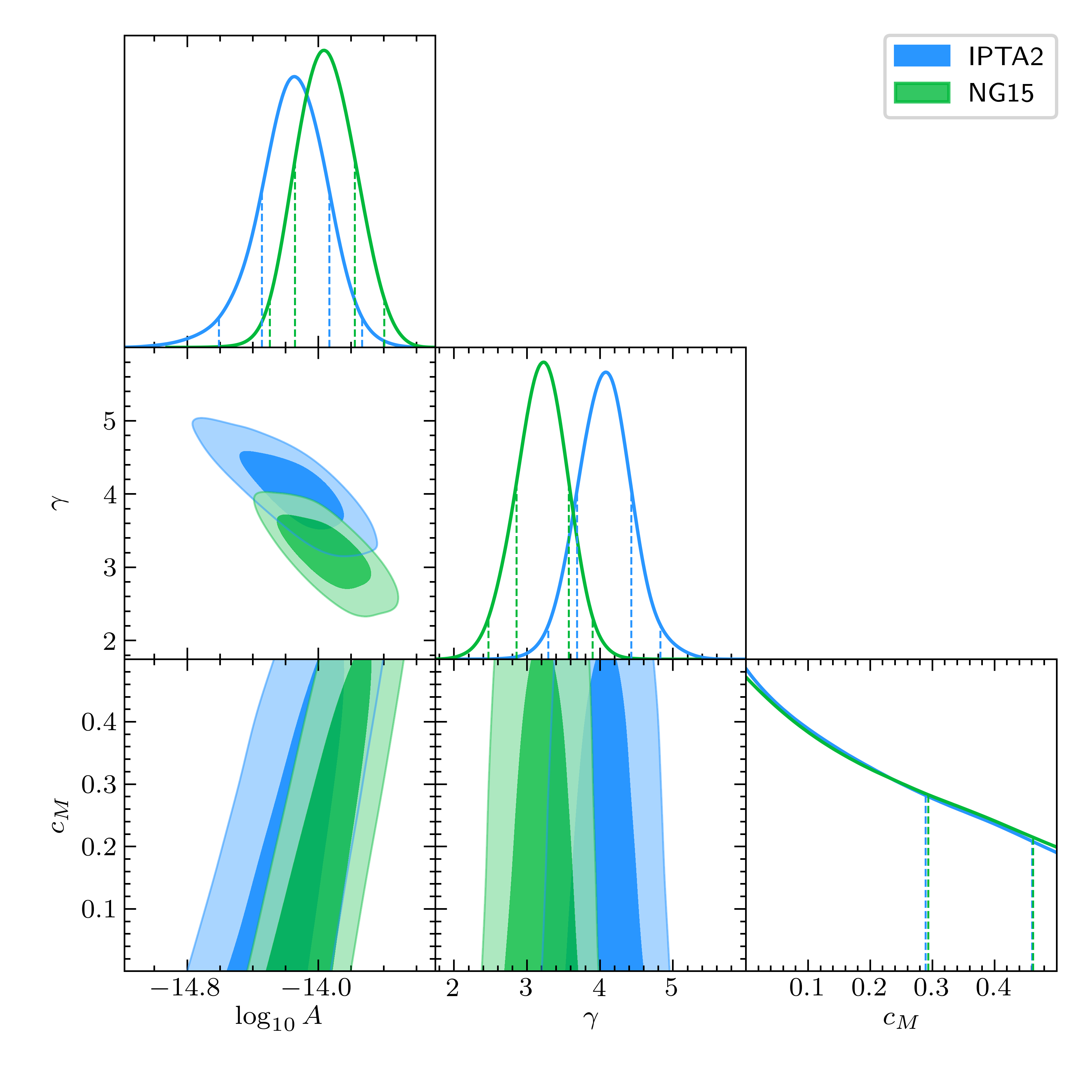}}
    	\caption{The posterior plots and marginal posteriors of no slip model parameters for \emph{parameterization II}. The contours show at 68\% and 95\%  confidence levels for International PTA second data release (IPTA2) in blue and NANOGrav 15-year data set (NG15) in green.}
    	\label{figfppar2}
    \end{figure}
    The results for \emph{parameterization I} using NG15 and IPTA2   are illustrated  in figure \ref{figfppar1}. The green lines show NG15 and blue lines show IPTA2. The results from both observations are consistent given the uncertainties on the parameters.
    We find that there exists a strong correlation between $\xi$ and $log_{10} A$. This degeneracy will hinder our ability to constrain them precisely.
    The constraints are summarized in table \ref{compare1}. To compare our study with other observations in table \ref{compare1}, we have provided results from our previous study in \cite{Allahyari:2021enz} where we generated  three mock standard sirens catalogs based on the merger of massive black hole binaries which are expected to be observed with LISA \cite{LISA:2017pwj,Baker:2019nia}. We combined the mock catalog and CMB observations  to constrain the parameters of no slip model. The second row in table~\ref{compare1}, shows the precision of different studies.

    The results for \emph{parameterization II} are provided in figure~\ref{figfppar2}. The green lines show NG15 and blue lines show IPTA2. We find  that
    $c_M$ and $log_{10} A$ are also correlated. Increasing $c_M$ has the effect of increasing the amplitude. In this case we only find an upper bound as $c_M<0.2$.
    We compare the results for $c_M$ in table \ref{compare2}  with results from the CMB study \cite{Brush:2018dhg}. It is seen that the current PTAs data can not perform better in constraining the model. 
    
    \begin{table*}
    	\centering
    	\begin{tabular}{ccc|ccc}
    		\hline
    		Parameter &  NG15 & IPTA2 & Pop III + CMB & Delay + CMB & No Delay + CMB \\
    		\hline 
    		
    		$\xi$ & $0.98^{+0.40}_{-0.84}$ & $0.97^{+0.44}_{-0.82}$ & $ 1.016^{+0.069}_{-0.083} $ & $1.13^{+0.14}_{-0.34}$ & $1.037^{+0.076}_{-0.17}$ \\
    		
    		$|\frac{\Delta \xi}{\xi}|$ & $1.27$ & $1.30$ & $0.15$ & $0.43$ & $0.24$ \\
    		
    		$log_{10}A$ & $-14.23^{+0.45}_{-0.22}$ & $-14.44^{+0.49}_{-0.23}$&  $-$  &  $-$  &  $-$   \\
    		
    		$\gamma$ & $3.21\pm 0.36$ & $4.06\pm 0.38$& $-$  &  $-$ & $-$   \\
    		\hline
    		
    	\end{tabular}
    	\caption{Constraints at 68\% confidence level from (NG15), (IPTA2), (Pop III + CMB), (Delay + CMB) and (No Delay + CMB) \cite{Allahyari:2021enz}, and their relative errors for \emph{parameterization I}.}
    	\label{compare1}
    \end{table*}

    \begin{table*}
    	\centering
    	\begin{tabular}{ccc}
    		\hline
    		Parameter &  NG15 & IPTA2  \\
    		\hline \\
    		
    		$c_M$ & $<0.293$ & $<0.289$  \\ \\
    		
    		$log_{10}A$ & $-13.95\pm 0.18$ & $-14.16^{+0.23}_{-0.18}$  \\ \\
    		
    		$\gamma$ & $3.20\pm 0.36$ & $4.07\pm 0.39$  \\
    		\hline
    		
    	\end{tabular}
    	\caption{Constraints at 68\% confidence level from NG15 and IPTA2 for \emph{parameterization II}.}
    	\label{compare2}
    \end{table*}

\begin{figure}[htbp!]
    	\centerline{\includegraphics[scale=1.2]{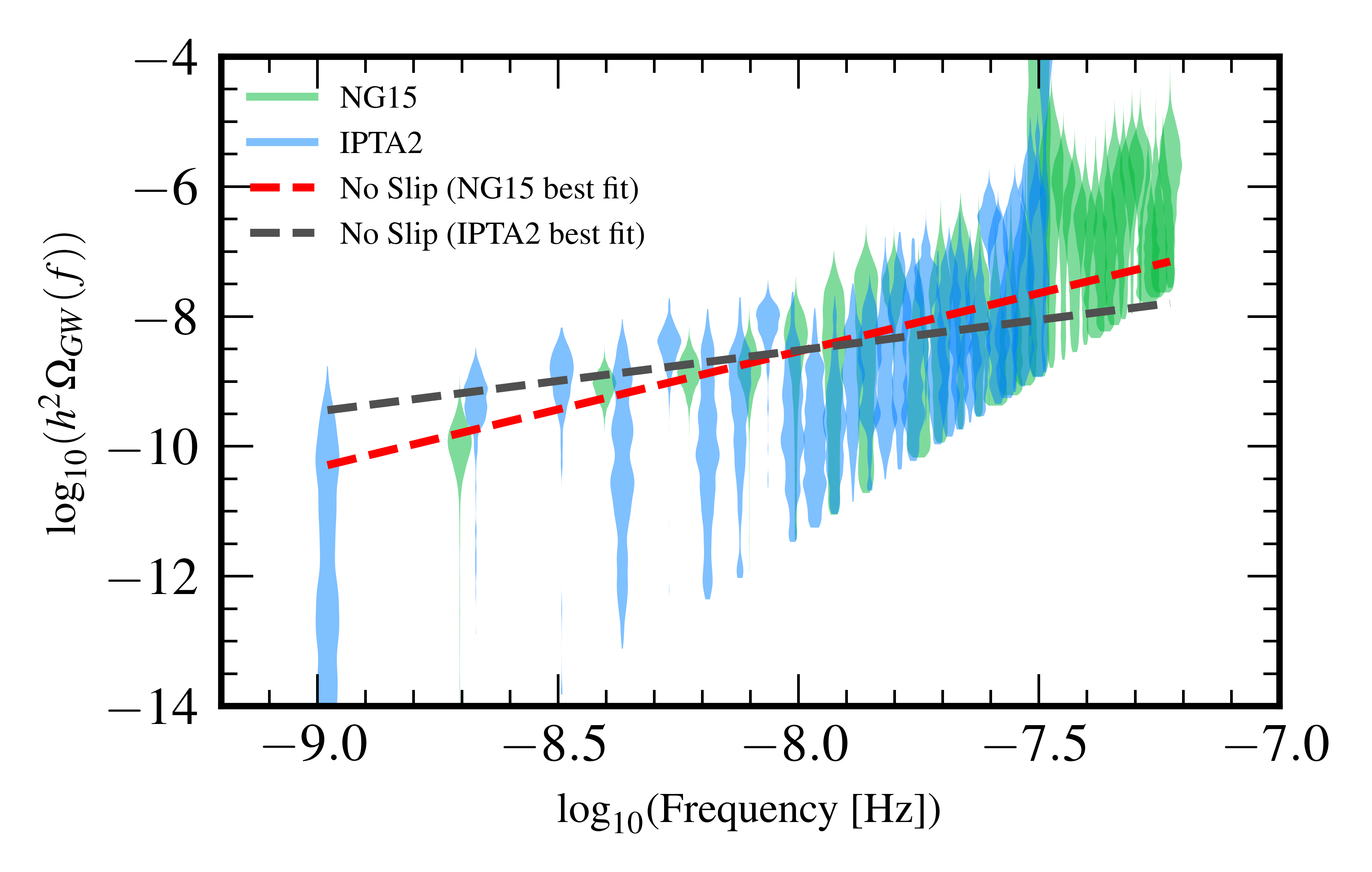}}
    	\caption{The current spectral energy density of GWs as a function of frequency in logarithmic scales for no slip model for \textbf{parameterization I}. The violin plots show NANOGrav 15-year data set (NG15) and International PTA second data release (IPTA2).    
     }
    	\label{omg1}
    \end{figure}

\begin{figure}[htbp!]
    	\centerline{\includegraphics[scale=1.2]{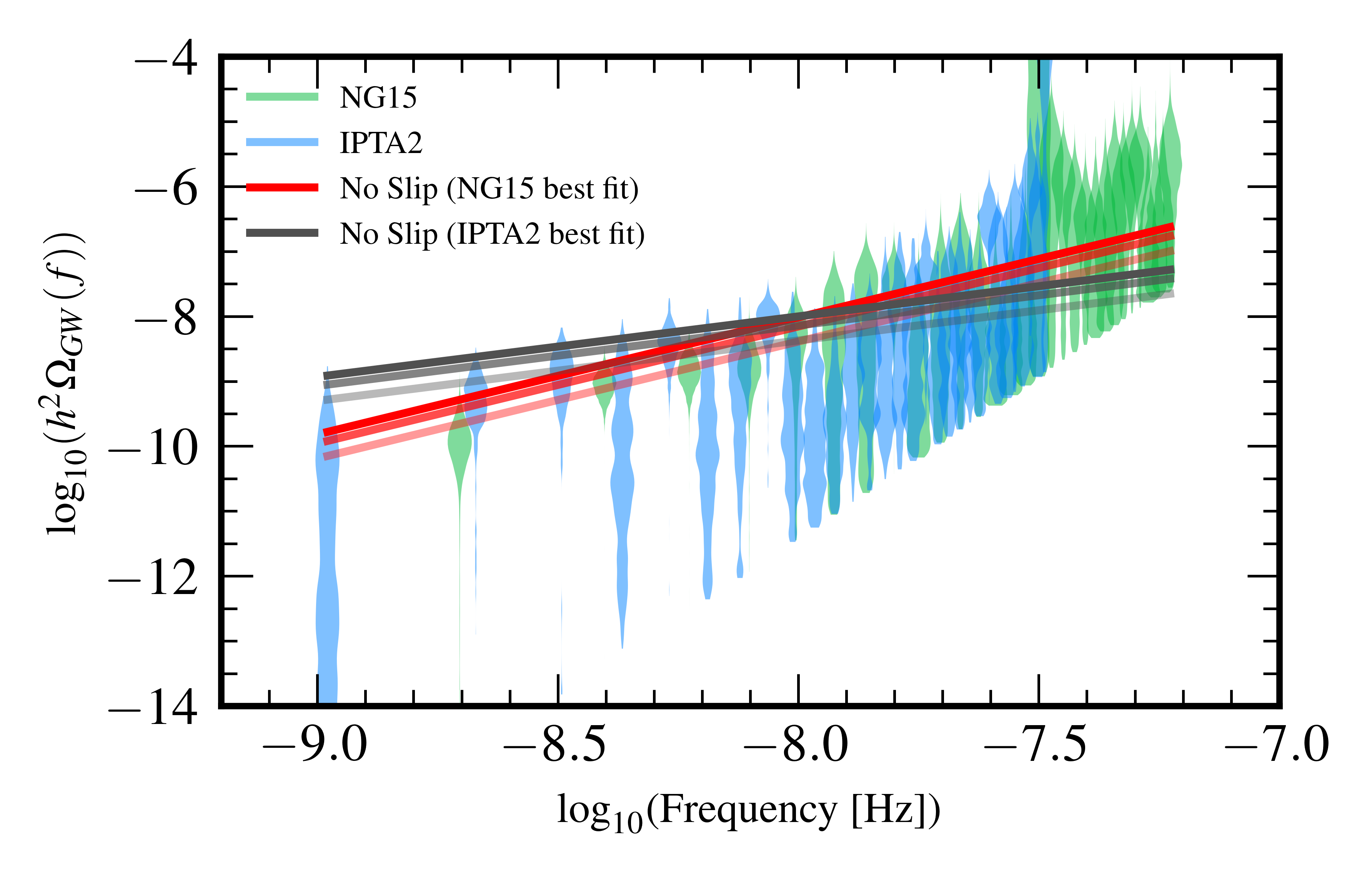}}
    	\caption{The current spectral energy density of GWs as a function of frequency in logarithmic scales for no slip model  for \textbf{parameterization II}. Respectively, dark red (gray) to pale red (gray) refers to $c_M=0.01$, $c_M=0.1$, $c_M=0.25$. The violin plots show NANOGrav 15-year data set (NG15) and International PTA second data release (IPTA2).    
    	}
    	\label{omg2}
    \end{figure}

    In figure \ref{omg1} and \ref{omg2}, for concreteness, we show $h^2\Omega_{GW}$ as a function of frequency in logarithmic scales using the best fit of values of the parameters obtained from NG15 and IPTA2 for \emph{parameterization I} and \emph{parameterization II}, respectively. $h$ is the reduced Hubble constant. The violin plots are from NG 15-year and IPTA second data release. 
    
\section{Summary and discussion} \label{sec5}
In this work, we constrained a modified gravity model in PTA scales.
We studied no slip model as a subclass of Horndeski models. In this model GWs have a different damping term compared to the standard general relativity case. The speed of GWs is equal to the speed of light in this model. Using two different parametrizations, we derived approximate transfer functions for GWs in this model. Moreover, we used this transfer function to find $\Omega_{GW}$ in our model.

We constrained the model parameters  by using NANOGrav and IPTA second data release. Two parameters $\xi$ and $c_M$ are correlated and degenerate with other parameters, so it is hard to constrain them with percent level accuracy. 

In our work, we only considered PTAs. One extension is to use multi probes of GWs by combining the existing GW detectors or future detectors like Einstein Telescope to estimate the constraints on the parameters~\cite{Jiang:2023qht, Liu:2022umx}. Additionally,  more thorough investigations could take into account various effects which could contaminate PTA signals.  Effects like  the interstellar medium, the solar wind and solar system ephemeris errors contribute to the noise~ \cite{Reardon:2023zen}.

\clearpage

\bibliographystyle{JHEP}
\bibliography{ref}

\end{document}